# N-continuous OFDM: System Optimization and Performance Analysis

Peng Wei, Lilin Dan, Yue Xiao, Wei Xiang, Shaoqian Li
National Key Laboratory of Science and Technology on Communications
University of Electronic Science and Technology of China
Chengdu, China
wpwwwhttp@163.com

*Abstract*—*N*-continuous orthogonal frequency division multiplexing (NC-OFDM) is a promising technique to achieve significant sidelobe suppression of baseband OFDM signals. However, the high complexity limits its application. Based on conventional NC-OFDM, in this paper, a new technique, called time-domain *N*-continuous OFDM (TD-NC-OFDM), is proposed to transfer the original frequency-domain processing to the time domain, by the linear combination of a novel basis set to smooth the consecutive OFDM symbols and their high-order derivatives. We prove that TD-NC-OFDM is an equivalent to conventional one while consuming much lower complexity. Furthermore, via the time-domain structure, a closed-form spectral expression of NC-OFDM signals is derived and a compact upper bound of sidelobe decaying is derived. This paper also investigates the impact of the TD-NC-OFDM technique on received signal-to-interference-plus-noise ratio (SINR) and provides a closed-form analytical expression. Theoretical analyses and simulation results show that TD-NC-OFDM can prohibitively suppress the sidelobe with much lower complexity.

## I. INTRODUCTION

Orthogonal frequency division multiplexing (OFDM) has been the most popular multicarrier transmission technique in wireless communications [1], [2] for its rapid data transmission and effective robustness against the inter-symbol interference (ISI) over frequency selective wireless channel. However, in rectangularly pulsed OFDM system, the signal possesses discontinuous pulse edge and exhibits larger spectrum sidelobe. Thus, power leakage, which is also known as out-of-band power emission, generated by high sidelobes, causes severe interference to adjacent channels [3]-[5].

Various methods have been proposed for sidelobe suppression. The windowing techniques [6], [7] utilize varying window functions to suppress the sidelobes. However, a largely extended guard interval for avoiding the signal distortion will reduce the system throughput. Furthermore, lots of methods [8]-[32] were presented for sidelobe suppression. Among them, *adaptive symbol transition* [8], *subcarrier weighting* [9], and *constellation expansion* (CE) [10], [11] cannot obtain appreciated spectrum roll-off. Subsequently, *interference cancellation* [12]-[16], *precoding* [17]-[24] and *N-continuous* techniques [25]-[32] have received wide attention recently.

In interference cancellation methods, the high signal sidelobes are minimized by optimizing the frequency-domain signal at cancellation subcarriers. However, cancellation carriers (CCs) [12]-[14] waste extra power with signal-to-noise ratio (SNR) loss, and active interference cancellation (AIC) methods [15], [16] cause serious ISI during multiple symbols with large-scale matrices operations. Additionally, the usage of reserved subcarrier in this class of methods will incur peak-to-average-power ratio (PAPR) growth. On the other hand, precoding methods [17]-[24] utilize frequency-domain precoders with optimal matrices to prohibitively mitigate power leakage. However, their large-scale matrices operations introduce heavy computational complexity.

*N*-continuous techniques smooth the amplitude and phase of adjacent OFDM symbols and their derivatives to achieve rapid sidelobe decaying. Conventional *N*-continuous OFDM (NC-OFDM) [26] makes OFDM signal *N*-continuous at the cost of high complexity caused by large-scale matrices operations. The reserved subcarrier based on NC-OFDM [25] also results in degradation of power efficiency as CCs with PAPR growth. Aiming at optimizing the frequency-domain precoder in NC-OFDM, Beek et al. further presented three methods in [27]-[29]. Unfortunately, the memoryless scheme [27], which makes the edge of OFDM symbols and their derivatives zero, will degrade the PAPR performance. According to the technique in [28] which nulls the spectrum at several chosen frequencies, the precoder in [29] enhances BER performance with data rate loss. On the other hand, to obtain low-complexity signal recovery for reception in NC-OFDM, multiple researches [30]-[32] had been proposed. However, the technique in [30], which develops NC-OFDM with selective mapping (SLM), causes significant increase in complexity during searching the optimal data sequence with minimum interference. Ref. [31] also causes high complexity in order to calculate the solution on cancellation tones. The precoder [32] combines two concepts of the precoders in [26] and [29] at the cost of data loss.

In this paper, we optimize the design and analyze the performance of NC-OFDM systems. On one hand, a new technique, referred as time-domain *N*-continuous OFDM (TD-NC-OFDM), is proposed to make OFDM signal *N*-continuous with much low complexity. In contrast to the frequency-domain precoder in NC-OFDM as described in current literature [25]-[32], TD-NC-OFDM models an optimization by adding a smooth signal in the time domain to smooth the OFDM signal. Then, the smooth signal is constructed by linear combination of a proposed novel basis set. We prove the existence of the basis set and its effect on smooth signal reconstruction, by analyzing the reconstruction of the derivatives of discrete-time OFDM signals. Moreover, we consider the generation of the basis set based on system configurations, and subsequently give a simple way for the calculation of the coordinates corresponding to the basis set, for generating the smooth signal. At last, we prove that TD-NC-OFDM is equivalent to NC-OFDM, which implies that TD-NC-OFDM has identical sidelobe suppression performance as conventional NC-OFDM.

On the other hand, based on the developed time-domain structure, we consider the performance analysis of NC-OFDM in terms of spectrum roll-off, complexity and signal-to-interference-plus-noise ratio (SINR). Firstly, via the time-domain continuity, which is measured by the number of continuous derivatives, a closed-form spectral expression of NC-OFDM signals is derived. The spectral analysis gives a compact upper bound of sidelobe which decays as at least $f^{-V-2}$, where $V$ is the maximum derivative order (MDO). Secondly, a brief complexity comparison between NC-OFDM and TD-NC-OFDM is given, which shows that TD-NC-OFDM is with significant complexity reduction. Finally, we analyze the SINR of the received OFDM signal interfered by the smooth signal, and give an exact expression for evaluating the system performance of NC-OFDM.

The rest of the paper is organized as follows. In section II, the problem of power leakage is formulated, and the relationship between continuity and spectrum roll-off is analyzed. Section III summarizes the traditional NC-OFDM. Section IV proposes the model of TD-NC-OFDM, proves some properties of derivatives of OFDM signal, designs the smooth signal by the linear combination of the basis set, and gives the OFDM transmitter with TD-NC-OFDM. Furthermore, we prove the equivalence between TD-NC-OFDM and NC-OFDM. In section V, the effect of TD-NC-OFDM on sidelobe decaying and complexity is analyzed as well as SINR. Finally, the conclusion is taken in section VI.

*Notations*: Boldfaced lowercase and uppercase letters are used to represent column vectors and matrices, respectively. $\{\mathbf{a}\}_n$ and $\{\mathbf{A}\}_{m,n}$ are the $n$th element of vector $\mathbf{a}$ and the element in the $m$th row and $n$th column of matrix $\mathbf{A}$, respectively. The $M \times M$ identity matrix and $M \times N$ zero matrix are denoted by $\mathbf{I}_M$ and $\mathbf{0}_{M \times N}$, respectively. $|\cdot|$ stands for absolute value. The trace and expectation are represented by $tr\{\cdot\}$, $E\{\cdot\}$, respectively. $\mathbf{A}^T$, $\mathbf{A}^*$, $\mathbf{A}^H$ and $\mathbf{A}^{-1}$ separately denote transposition, conjugate, conjugate transposition, inverse of matrix $\mathbf{A}$. $(\cdot)^{(v)}$ is defined by the $v$th-order derivative.

## II. SYSTEM ASPECT

In this section, a brief description of the OFDM system, the power leakage problem as well as the relationship between continuity and sidelobe decaying, is given. In a baseband OFDM system, the bit stream is first modulated to create a complex-valued data vector $\mathbf{X}_i = [X_{i,k_0}, X_{i,k_1}, \ldots, X_{i,k_{K-1}}]^T$ drawn from an $M$-ray constellation, such as $M$-ray phase-shift keying ($M$-PSK) or $M$-ray quadrature amplitude modulation ($M$-QAM). The complex-valued data are mapped to $K$ subcarriers with index set $\mathcal{K} = \{k_0, k_1, \ldots, k_{K-1}\}$. An OFDM signal is formed by summing all the $K$-modulated orthogonal subcarriers with equal frequency spacing $\Delta f = 1/T_s$, where $T_s$ is the duration of OFDM symbol. The $i$th OFDM time-domain symbol, assuming a rectangular time-domain window $R(t)$ [20], is expressed as

$$y_i(t) = \sum_{m=0}^{K-1} X_{i,k_m} e^{j2\pi k_m \Delta f t}, \quad -T_{cp} \leq t < T_s \quad (1)$$

in which $T_{cp}$ is the cyclic prefix (CP) duration. Then, in the time range of $(-\infty, +\infty)$, the total transmitted OFDM signal $s(t)$ can be written by

$$s(t) = \sum_{i=-\infty}^{+\infty} y_i \left( t - i(T_s + T_{cp}) \right). \quad (2)$$

By transforming the windowed time-domain subcarrier waveform into the frequency domain, the frequency-domain representation of the $k_m$th subcarrier, $C_{i,k_m}(f)$, is given by [20]

$$\begin{aligned} C_{i,k_m}(f) &= \frac{1}{T_s + T_{cp}} \int_{-\infty}^{+\infty} R(t) X_{i,k_m} e^{j2\pi \frac{k_m}{T_s} t} e^{-j2\pi f t} dt \\ &= X_{i,k_m} \operatorname{sinc}\left( f_m (1 + T_{cp}/T_s) \right) e^{j\pi f_m (1 - T_{cp}/T_s)} \end{aligned} \quad (3)$$

where $\operatorname{sinc}(x) \triangleq \sin(\pi x)/(\pi x)$ and $f_m = k_m - fT_s$. $C_{i,k_m}(f)$ also denotes the interference coefficient from the subcarrier $k_m$ to the frequency $f$ which is outside the allocated frequency band. Eq. (3) shows that the sidelobe of rectangularly pulsed OFDM signal decays as $f^{-1}$ when $f$ is large. The sum of the interference coefficients in all subcarriers results in power leakage. Thus, the power spectrum density (PSD) of $s(t)$ is expressed by [22] as

$$\begin{aligned} \psi(f) &= (T_s + T_{cp}) E \left\{ \left| \sum_{m=0}^{K-1} C_{i,k_m}(f) \right|^2 \right\} \\ &= (T_s + T_{cp}) E \left\{ \left| \sum_{m=0}^{K-1} X_{i,k_m} e^{j\pi f_m (1 - T_{cp}/T_s)} \right. \right. \\ &\quad \left. \left. \operatorname{sinc}\left( f_m (1 + T_{cp}/T_s) \right) \right|^2 \right\}. \end{aligned} \quad (4)$$

According to (3) and (4), the high sidelobe is due to the rectangular window $R(t)$ which causes the discontinuity between adjacent symbols. As mentioned in [33], the smoother a signal is, measured by the number of continuous derivatives it possesses, the faster its spectrum sidelobe dies away. Suppose that $s(t)$ and its $v$th-order derivative $s^{(v)}(t)$ are integrable and they approach zero as $t$ approaches $\pm\infty$, where $v \in \mathcal{U} = \{0,1,...,V\}$ with the MDO $V$. By extending the case in [33] to this assumption, the signal spectrum follows

$$S(f) = \frac{1}{(j2\pi f)^V} \int_{-\infty}^{+\infty} s^{(V)}(t) e^{-j2\pi ft} dt. \quad (5)$$

Instead of the slow sidelobe decaying as $f^{-1}$ in (3), the signal with its first $V$ derivatives continuous has rapidly decreasing spectrum sidelobe, which decays as at least $f^{-V}$ when $f$ is large. It is inferred that the continuity of OFDM signal, as measured by the number of continuous derivatives it possesses, dominates the rate at which its spectrum sidelobe decays. Consequently, the continuity enhancement of time-domain OFDM signal can achieve notable power leakage reduction.

### III. N-CONTINUOUS OFDM

Following the continuity criterion, Beek et al. modeled a signal format to improve the continuity by smoothing the OFDM signal and its high-order derivatives [26]. N-continuous OFDM (NC-OFDM) introduces a frequency-domain precoder to make the time-domain OFDM signal N-continuous. After smoothing two consecutive OFDM symbol by the precoder, the $i$th N-continuous time-domain OFDM symbol $\bar{y}_i(t)$ and its first $V$ derivatives satisfy

$$\bar{y}_i^{(v)}(t)|_{t=-T_{cp}} = \bar{y}_{i-1}^{(v)}(t)|_{t=T_s}. \quad (6)$$

Based on (1) and (6), the precoded process can be summarized as

$$\begin{cases} \bar{\mathbf{X}}_i = \mathbf{X}_0, & i = 0 \\ \bar{\mathbf{X}}_i = (\mathbf{I}_K - \mathbf{P})\mathbf{X}_i + \mathbf{P}\mathbf{\Phi}^H \bar{\mathbf{X}}_{i-1}, & i > 0 \end{cases} \quad (7)$$

where $\mathbf{P} = \mathbf{\Phi}^H \mathbf{A}^H (\mathbf{A}\mathbf{A}^H)^{-1} \mathbf{A}\mathbf{\Phi}$, $\mathbf{\Phi} = diag(e^{j\varphi k_0}, e^{j\varphi k_1},...,e^{j\varphi k_{K-1}})$, $\varphi = -2\pi T_{cp}/T_s$ and $\{\mathbf{A}\}_{v+1,m+1} = k_m^v$. Fig. 1 depicts the spectrally precoded NC-OFDM transmitter. The $i$th frequency-domain data vector $\mathbf{X}_i$ is first precoded. The precoded data vector $\bar{\mathbf{X}}_i = [\bar{X}_{i,k_0}, \bar{X}_{i,k_1},\cdots,\bar{X}_{i,k_{K-1}}]^T$ is then fed into inverse fast Fourier transform (IFFT), and finally CP is added to generate the transmission signal.

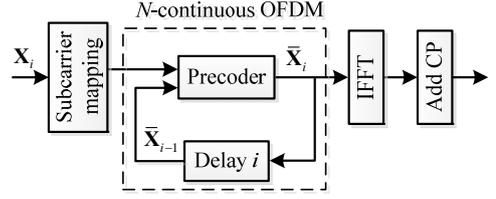

Fig. 1. The block diagram of N-continuous OFDM transmitter.

NC-OFDM can achieve significant suppression of out-of-band power leakage. However, its N-continuous precoder significantly increases the complexity of the transmitter, since $\mathbf{P}$ and $\mathbf{\Phi}$ are both $K \times K$ matrices, in which the complexity of the large-scale matrices operation is proportional to the square of number of subcarriers, i.e. $o(K^2)$. In addition, it is inconvenient to analyze the sidelobe decaying of N-continuous signal $\bar{\mathbf{X}}_i$ precoded in the frequency domain.

### IV. TIME-DOMAIN N-CONTINUOUS OFDM

For solving the problem of high complexity and offering a model easy for analyzing, in this section, a time-domain smooth signal generation scheme is proposed. TD-NC-OFDM is also based on the continuity enhancement of high-order derivatives, and therefore, makes OFDM signal N-continuous, by adding a time-domain smooth signal. Furthermore, it obtains reduced-complexity transmitter and its time-domain expression is convenient to analyze as specified in section V.

This section is organized as follows. In section IV-A, we first propose the time-domain model of TD-NC-OFDM by introducing a smooth signal to eliminate the discontinuities between adjacent OFDM symbols and their high-order derivatives. To design the smooth signal, in section IV-B, we examine the time-domain reconstruction of continuous-time signal and its derivatives, and extend the analysis to the discrete-time multicarrier signals and their derivatives. Then, based on the property of multicarrier signal reconstruction in section IV-B, the smooth signal is designed by the linear combination of a novel basis set in section IV-C, which subsequently specifies the new basis set and its corresponding coordinates calculation. From the design of the smooth signal, section IV-D presents a low-complexity OFDM transmitter. Furthermore, in section IV-E, the equivalence between NC-OFDM and TD-NC-OFDM in signal processing is proved.

*A. System Model*

For convenient illustration of TD-NC-OFDM technique in digital signal processing, in the rest of the paper, a discrete-time representation of the OFDM signal will be used, which is oversampled by time-domain sampling interval $T_{samp} = T_s/N$ with subcarrier index set $\mathcal{K}_0 = \{-\frac{K}{2},...,\frac{K}{2}-1\}$ and CP duration of length $N_{cp}$ in samples.

Similar to NC-OFDM, TD-NC-OFDM is also based on adding a smooth signal, but its processing is performed in the time domain. After the addition of the $i$th smooth signal $w_i(n)$

to OFDM symbol $y_i(n)$, the $i$th smoothed time-domain symbol $\bar{y}_i(n)$ is written by

$$\bar{y}_i(n) = y_i(n) + w_i(n) \tag{8}$$

where $n \in \mathcal{N} = \{-N_{cp},\ldots,0,\ldots,N-1\}$ is the sample index. Except the discontinuities at the edge of each OFDM symbol, the OFDM signal possesses derivatives of all orders. Thus, the smooth signal works on a pair of adjacent OFDM symbols to mitigate the unique discontinuity at the adjacent point between two consecutive OFDM symbols. According to the time-domain operation in (8), the smooth signal is distributed in the whole symbol. Hence, the discontinuity is eliminated by making it bicontinuous at adjacent point. In other words, the smooth signal is required to consider the effect of both the current and former OFDM symbol. Therefore, to smooth two consecutive symbols, as measured by the continuity enhancement corresponding to adjacent OFDM symbols and their first $V$ derivatives, the smooth signal and its first $V$ derivatives should follow that

$$w_i^{(v)}(n)\big|_{n=-N_{cp}} = \bar{y}_{i-1}^{(v)}(n)\big|_{n=N} - y_i^{(v)}(n)\big|_{n=-N_{cp}}. \tag{9}$$

In fact, model (9) can also make the OFDM signal $N$-continuous at the adjacent points, similar to traditional NC-OFDM. Therefore, it is possible to make OFDM signal $N$-continuous by time-domain processing.

### B. Analysis of Time-domain Signal and Its High-order Derivatives

In TD-NC-OFDM, the key design issue is to generate the smooth signal in the time domain. Firstly, we examine the time-domain reconstruction of continuous-time signal and its high-order derivatives. Then, a generalized reconstruction of discrete-time multicarrier signals and their high-order derivatives is established. Furthermore, a basis set drawn from a basis function and its derivatives is presented.

Using the mathematical argument of the interpolation for continuous-time signals in [34], we state and prove the important results on reconstruction of derivatives of time-domain signal precisely as *Theorem 1* and *Proposition 1*. In *Theorem 1*, we show that all derivatives of time-domain signal $s(t)$ exist in the time range $t \in (-\infty, +\infty)$.

*Theorem 1*: If the $v$th-order derivative of continuous-time signal $s^{(v)}(t)$ exists ($v \in \mathcal{U}$), then it can be reconstructed by the linear interpolation of the signal value at sampling time and the $v$th-order derivative of the *sinc* function.

The proof is shown in Appendix A.

A time-domain reconstruction of the oversampled multicarrier signal is established in *Proposition 1* using *Theorem 1*. We first make the assumption that the first $V$ derivatives of the oversampled signal $s(n)$ all exist in discrete time.

*Proposition 1*: In a baseband multicarrier system with CP, if the $v$th-order derivative of the oversampled signal $s^{(v)}(n)$ exists ($v \in \mathcal{U}$), then the $v$th-order derivative of the $i$th time-domain symbol $y_i^{(v)}(n)$ can be expressed as the cyclic convolution of the $i$th symbol value $\tilde{y}_i(n)$ at non-oversampling point and the $v$th-order derivative $f^{(v)}(n)$ of the basis function $f(n)$, expressed as

$$y_i^{(v)}(n) = \tilde{y}_i(n) \circledast f^{(v)}(n), \tag{10}$$

where the symbol $\circledast$ implies cyclic convolution and the $v$th-order derivative $f^{(v)}(n)$ is designed by (47) based on subcarrier location $\mathcal{K}$, number of symbol samples and CP length.

The proof of *Proposition 1* is illustrated in Appendix B.

It is implied in *Proposition 1* that the derivative set of time-domain multicarrier symbol $\{\mathbf{y}_i, \mathbf{y}_i^{(1)},\ldots,\mathbf{y}_i^{(V)}\}$ with dimension $V+1$ can be seen as the projection of the original symbol vector $\mathbf{y}_i$ onto a basis set

$$\mathcal{O} = \{\mathbf{q}_v \mid \mathbf{q}_v = [f^{(v)}(-N_{cp}), f^{(v)}(-N_{cp}+1),\ldots,f^{(v)}(N-1)]^T\}. \tag{11}$$

The proposition is apparently appropriate to OFDM system. Therefore, according to the linear property in (10) and the proposed basis set $\mathcal{O}$, the smooth signal based on a novel basis set is designed as follows.

### C. Design of the Smooth Signal Based on A Novel Basis Set

Inspired by the construction of the time-domain OFDM signal as described in *Proposition 1*, the smooth signal is designed by

$$w_i(n) = \sum_{v=0}^{V} b_{i,v} f^{(v)}(n) \tag{12}$$

where the linear combination of the smooth signal is concerned with the coordinate $b_{i,v} \in \mathbf{b}_i = [b_{i,0}, b_{i,1},\ldots,b_{i,V}]^T$ which will be calculated later, and a new basis set

$$\mathcal{Q} = \{\mathbf{q}_{\tilde{v}} \mid \mathbf{q}_{\tilde{v}} = [f^{(\tilde{v})}(-N_{cp}),\ldots,f^{(\tilde{v})}(N-1)]^T, \\ \tilde{v} \in \{0,1,\ldots,2V\}\}. \tag{13}$$

In the basis set $\mathcal{Q}$, the basis vectors $\mathbf{q}_{\tilde{v}}$ is related to subcarrier location $\mathcal{K}_0$, number of symbol samples $N$ and CP length $N_{cp}$. In fact, the smooth signal can be considered as a set projection onto the subspace consisting of the first $V+1$ basis vectors in the basis set $\mathcal{Q}$ with the coordinates $b_{i,v}$ ($v \in \mathcal{U}$).

The smooth signal is composed of two parts, as the basis set $\mathcal{Q}$ and its corresponding coordinates $b_{i,v}$. In the following, how the basis set and the coordinates are obtained is specified, respectively. As the set $\mathcal{O}$ analyzed in Appendix B, in the case of the number of symbol samples $N$, CP length $N_{cp}$ and the index set $\mathcal{K}_0$, the element $f^{(\tilde{v})}(n)$ of the basis vector $\mathbf{q}_{\tilde{v}}$ is calculated by

$$f^{(\tilde{v})}(n) = \begin{cases} \dfrac{\sin\left(\dfrac{\pi K}{N}(n+N_{cp})\right)}{N\sin\left(\dfrac{\pi}{N}(n+N_{cp})\right)} e^{j\frac{\pi}{N}(1-2K)(n+N_{cp})} = f(n), & \tilde{v}=0 \\ \dfrac{1}{N}(j\dfrac{2\pi}{N})^{\tilde{v}} \sum_{k_m \in \mathcal{K}_0} k_m^{\tilde{v}} e^{-j\varphi k_m} e^{j2\pi\frac{k_m}{N}n}, & \tilde{v}>0 \end{cases}.$$
(14)

In addition, $f^{(\tilde{v})}(n)$ can be derived from the $\tilde{v}$ th-order derivative of $f(n)$ by

$$f^{(\tilde{v})}(n) = \frac{d^{\tilde{v}}}{dn^{\tilde{v}}} f(n). \tag{15}$$

Therefore, the basis set $\mathcal{Q}$ can be easily generated by (14) or (15) in advance.

With the basis set $\mathcal{Q}$, if the coordinates are resolvable, the generation of the smooth signal in (12) can be expected. In the following, we will show the calculation of the coordinates.

For the convenience of representation, Eq. (18) is rewritten by the matrix form as

$$\mathbf{w}_i = \mathbf{Q}_f \mathbf{b}_i \tag{16}$$

where the basis matrix $\mathbf{Q}_f = [\mathbf{q}_0 \ \mathbf{q}_1 \ \cdots \ \mathbf{q}_V]$ consists of the first $(V+1)$ basis vectors in the basis set $\mathcal{Q}$.

Substituting (8) and (16) into (9), we obtain

$$\mathbf{P}_f \mathbf{b}_i = \Delta \mathbf{y}_i. \tag{17}$$

Because $\mathbf{P}_f$ is a $(V+1) \times (V+1)$ symmetric matrix without linear correlation among its rows or columns, which is given as

$$\mathbf{P}_f = \begin{bmatrix} f(-N_{cp}) & f^{(1)}(-N_{cp}) & \cdots & f^{(V)}(-N_{cp}) \\ f^{(1)}(-N_{cp}) & f^{(2)}(-N_{cp}) & \cdots & f^{(V+1)}(-N_{cp}) \\ \vdots & \vdots & & \vdots \\ f^{(V)}(-N_{cp}) & f^{(V+1)}(-N_{cp}) & \cdots & f^{(2V)}(-N_{cp}) \end{bmatrix},$$

its inverse $\mathbf{P}_f^{-1}$ exists. Therefore, it is concluded that the coordinates exist and are solvable. Vector $\Delta \mathbf{y}_i$ denotes the differences of two consecutive OFDM symbols including their first $V$ derivatives at the adjacent point, expressed as

$$\Delta \mathbf{y}_i = \begin{bmatrix} \bar{y}_{i-1}(N) - y_i(-N_{cp}) \\ \bar{y}_{i-1}^{(1)}(N) - y_i^{(1)}(-N_{cp}) \\ \vdots \\ \bar{y}_{i-1}^{(V)}(N) - y_i^{(V)}(-N_{cp}) \end{bmatrix}.$$

Via the following matrices operations, the vector $\Delta \mathbf{y}_i$ can be calculated as

$$\Delta \mathbf{y}_i = \begin{bmatrix} \bar{y}_{i-1}(N) - y_i(-N_{cp}) \\ \mathbf{P}_1 \bar{\mathbf{y}}_{i-1} - \mathbf{P}_2 \mathbf{X}_i \end{bmatrix} \tag{18}$$

where the element $\{\mathbf{P}_1\}_{v,n+1}$ is $\dfrac{1}{N} \sum_{k_m \in \mathcal{K}_0} (\dfrac{j2\pi}{N} k_m)^v e^{-j\frac{2\pi k_m n}{N}}$ and the element $\{\mathbf{P}_2\}_{v,m+1}$ is $\dfrac{1}{N}(\dfrac{j2\pi}{N} k_m)^v e^{j\varphi k_m}$ for $v \neq 0$. From (17) and (18), the coordinates are given by

$$\mathbf{b}_i = \mathbf{P}_f^{-1} \begin{bmatrix} \bar{y}_{i-1}(N) - y_i(-N_{cp}) \\ \mathbf{P}_1 \bar{\mathbf{y}}_{i-1} - \mathbf{P}_2 \mathbf{X}_i \end{bmatrix} \tag{19}$$

where the complexity of the operations of matrices $\mathbf{P}_1$ and $\mathbf{P}_2$ has been significantly reduced compared to NC-OFDM. However, $\mathbf{P}_1$ is still a large-scale matrix when the number of symbol samples is large. Thus, by reducing the scale of the matrix $\mathbf{P}_1$, the calculation of the coordinate vector $\mathbf{b}_i$ is further optimized as

$$\mathbf{b}_i = \mathbf{P}_f^{-1} \begin{bmatrix} \bar{y}_{i-1}(N) - y_i(-N_{cp}) \\ \tilde{\mathbf{P}}_1 \mathbf{X}_{i-1} + \mathbf{P}_V \mathbf{b}_{i-1} - \mathbf{P}_2 \mathbf{X}_i \end{bmatrix}, \tag{20}$$

where element $\{\tilde{\mathbf{P}}_1\}_{v,m+1}$ is $\dfrac{1}{N}(\dfrac{j2\pi}{N} k_m)^v$ for $v \neq 0$, and matrix $\mathbf{P}_V$ is used to calculate the derivative value of the $(i\text{-}1)$th smooth signal at the adjacent point, expressed as

$$\mathbf{P}_V = \begin{bmatrix} f^{(1)}(N) & f^{(2)}(N) & \cdots & f^{(V+1)}(N) \\ f^{(2)}(N) & f^{(3)}(N) & \cdots & f^{(V+2)}(N) \\ \vdots & \vdots & & \vdots \\ f^{(V)}(N) & f^{(V+1)}(N) & \cdots & f^{(2V)}(N) \end{bmatrix}.$$

### D. TD-NC-OFDM Transmitter Design

Fig. 2 depicts the block diagram of TD-NC-OFDM transmitter. Based on system configurations of subcarrier location $\mathcal{K}_0$, number of symbol samples $N$ and CP length $N_{cp}$, the basis set $\mathcal{Q}$ is determined in advance, and subsequently the basis matrix $\mathbf{Q}_f$ and parameter matrices including $\mathbf{P}_f^{-1}$, $\tilde{\mathbf{P}}_1$, $\mathbf{P}_V$ and $\mathbf{P}_2$ are preset. The original data vector $\mathbf{X}_i$ is first mapped onto the subcarrier index set $\mathcal{K}_0$ and then transformed to time-domain symbol $\mathbf{y}_i$ by IFFT. Lastly, TD-NC-OFDM operates in time domain with calculation of coordinate vector

$\mathbf{b}_i$, generation of the smooth signal vector $\mathbf{w}_i$, as well as the addition of the smooth signal to the time-domain OFDM symbol $\mathbf{y}_i$. After that, the smoothed time-domain OFDM symbol $\bar{\mathbf{y}}_i$ is expressed as

$$\bar{\mathbf{y}}_i = \mathbf{y}_i + \mathbf{Q}_f \mathbf{b}_i. \tag{21}$$

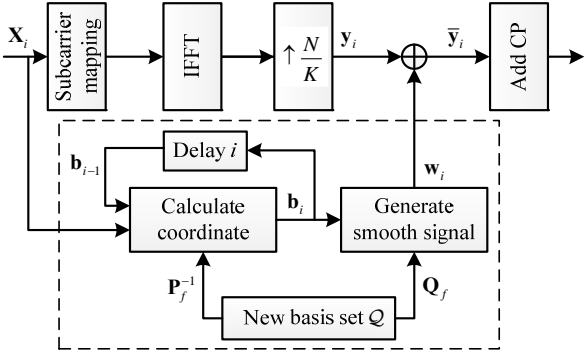

Fig. 2. The block diagram of TD-NC-OFDM Transmitter.

Note that for saving the memory, the first and last element of the last $V$ basis vectors in the set $\mathcal{Q}$ is stored, and only $V+1$ derivative values of the $(i-1)$th OFDM symbol at the adjacent points are stored after one symbol delay. According to the above design, TD-NC-OFDM is specified by the following steps:

(1) Initialization: $\mathbf{Q}_f$, $\mathbf{P}_f^{-1}$, $\tilde{\mathbf{P}}_1$, $\mathbf{P}_V$ and $\mathbf{P}_2$;

(2) For the first symbol at $i=0$, let $\bar{\mathbf{y}}_0 = \mathbf{y}_0$;

(3) When $i>0$, the coordinate vector $\mathbf{b}_i$ is calculated by (20), and the smoothed OFDM signal $\bar{\mathbf{y}}_i$ is generated by adding the smooth signal as (21).

*E. Equivalence of NC-OFDM and TD-NC-OFDM*

NC-OFDM and TD-NC-OFDM both focus on making OFDM signal $N$-continuous. However, they have different implementations as shown in Fig.1 and Fig. 2. NC-OFDM adopts a frequency-domain precoder and TD-NC-OFDM adds a smooth signal in the time domain. In this section, we prove the equivalence between NC-OFDM and TD-NC-OFDM.

In NC-OFDM [26], the precoder adds a frequency-domain noise $-\mathbf{P}\mathbf{X}_i + \mathbf{P}\boldsymbol{\Phi}^H \bar{\mathbf{X}}_{i-1}$ to get the precoded data $\bar{\mathbf{X}}$. TD-NC-OFDM also obtains the smoothed OFDM symbol $\bar{\mathbf{y}}_i$ by the addition of the smooth signal $\mathbf{w}_i$. Thus, the key problem is to figure out the relationship between $-\mathbf{P}\mathbf{X}_i + \mathbf{P}\boldsymbol{\Phi}^H \bar{\mathbf{X}}_{i-1}$ and $\mathbf{w}_i$. In the following, we will investigate their relationship by transforming the smooth signal in TD-NC-OFDM to the frequency domain. After Fourier transform, the frequency-domain expression of the smooth signal $\mathbf{w}_i$ in (17) is given by

$$\mathbf{W}_i = \mathbf{F}_f \mathbf{Q}_f \mathbf{b}_i \tag{22}$$

where $\mathbf{F}_f$ is the Fourier transform matrix corresponding to the index set $\mathcal{K}_0$, with $\frac{1}{N}\mathbf{F}_f \mathbf{F}_f^H = \mathbf{I}_K$. For the convenience of representation, in this subsection and Appendix C, both $\mathbf{P}_1$ and $\mathbf{P}_2$ include the row of $v=0$. From (19) and (22), we obtain

$$\mathbf{W}_i = \mathbf{F}_f \mathbf{Q}_f \mathbf{P}_f^{-1}(\mathbf{P}_1 \bar{\mathbf{y}}_{i-1} - \mathbf{P}_2 \mathbf{X}_i) \tag{23}$$

where the parameter matrices $\mathbf{Q}_f$, $\mathbf{P}_1$, $\mathbf{P}_2$ and $\mathbf{P}_f$ can be rewritten as

$$\mathbf{Q}_f = \frac{1}{N}\mathbf{F}_f^H \mathbf{B}_1, \tag{24}$$

$$\mathbf{P}_1 = \frac{1}{N}\mathbf{B}_2 \mathbf{F}_f, \tag{25}$$

$$\mathbf{P}_2 = \frac{1}{N}\mathbf{B}_2 \boldsymbol{\Phi} \tag{26}$$

and

$$\mathbf{P}_f = \frac{1}{N}\mathbf{B}_2 \mathbf{B}_2^T \tag{27}$$

where $\{\mathbf{B}_1\}_{m+1,v+1} = (j\frac{2\pi}{N}k_m)^v e^{-j\varphi k_m}$ and $\{\mathbf{B}_2\}_{v+1,m+1} = (j\frac{2\pi}{N}k_m)^v$, with $\mathbf{B}_1 = \boldsymbol{\Phi}^H \mathbf{B}_2^T$. Substituting (24)-(27) into (23), the frequency-domain smooth signal $\mathbf{W}_i$ can be rewritten by

$$\mathbf{W}_i = \mathbf{B}_1(\mathbf{B}_2 \mathbf{B}_2^T)^{-1}(\mathbf{B}_2 \bar{\mathbf{X}}_{i-1} - \mathbf{B}_2 \boldsymbol{\Phi} \mathbf{X}_i). \tag{28}$$

Compared to $-\mathbf{P}\mathbf{X}_i + \mathbf{P}\boldsymbol{\Phi}^H \bar{\mathbf{X}}_{i-1}$ in (7), if $\mathbf{P} = \mathbf{B}_1(\mathbf{B}_2\mathbf{B}_2^T)^{-1}\mathbf{B}_2$, the equivalence between NC-OFDM and TD-NC-OFDM will be proved.

From $\mathbf{A}$ and $\mathbf{B}_2$, we can obtain

$$\mathbf{B}_2 = \boldsymbol{\Phi}_1 \mathbf{A} \tag{29}$$

where $\boldsymbol{\Phi}_1 = diag(1, j\frac{2\pi}{N},...,(j\frac{2\pi}{N})^v)$. Substituting (29) into $\mathbf{B}_1(\mathbf{B}_2\mathbf{B}_2^T)^{-1}\mathbf{B}_2$, we obtain

$$\begin{aligned}\mathbf{B}_1(\mathbf{B}_2\mathbf{B}_2^T)^{-1}\mathbf{B}_2 &= \boldsymbol{\Phi}^H \mathbf{B}_2^T (\mathbf{B}_2\mathbf{B}_2^T)^{-1}\mathbf{B}_2 \\ &= \boldsymbol{\Phi}^H \mathbf{A}^T \boldsymbol{\Phi}_1 (\boldsymbol{\Phi}_1 \mathbf{A}\mathbf{A}^T \boldsymbol{\Phi}_1)^{-1}\boldsymbol{\Phi}_1 \mathbf{A} \\ &= \boldsymbol{\Phi}^H \mathbf{A}^T (\mathbf{A}\mathbf{A}^T)^{-1}\mathbf{A}.\end{aligned}$$

Because all elements in matrix **A** are real, we have $\mathbf{A}^T = \mathbf{A}^H$. Thus, we get that

$$\mathbf{B}_1(\mathbf{B}_2\mathbf{B}_2^T)^{-1}\mathbf{B}_2 = \mathbf{\Phi}^H \mathbf{A}^H (\mathbf{A}\mathbf{A}^H)^{-1}\mathbf{A} = \mathbf{P}. \quad (30)$$

Therefore, it is concluded that the frequency-domain expression of the smooth signal $\mathbf{W}_i$ is the *N*-continuous noise $-\mathbf{P}\mathbf{X}_i + \mathbf{P}\mathbf{\Phi}^H \overline{\mathbf{X}}_{i-1}$. Furthermore, we show that the proposed TD-NC-OFDM is the time-domain equivalent implementation of conventional NC-OFDM.

## V. ANALYSIS OF SPECTRUM, COMPLEXITY AND SINR

In this section, an analytical expression of the spectrum of NC-OFDM signal is developed. This expression, different from the one of the rectangularly pulsed OFDM signal, corresponds to an exact description of the smoothed time-domain OFDM signal. After deriving a closed-form expression for the spectrum of the smoothed OFDM symbol, we obtain a compact upper bound of spectrum power leakage of the smoothed OFDM signal. Then, we compare the computational complexity between NC-OFDM and TD-NC-OFDM, where TD-NC-OFDM achieves considerable complexity reduction. Lastly, by analyzing the average power of the smooth signal in time domain, we give an exact estimation of the received SINR.

### A. Spectral Analysis

The equivalence of NC-OFDM and TD-NC-OFDM has been proved in the former section. However, the conventional frequency-domain processing is inconvenient to analyze the sidelobe decaying. According to the spectrum analysis for the rectangularly pulsed signal in (4), the improved continuity of the smoothed OFDM signal will be removed by rectangular window so that the spectral trends out of band is not correctly demonstrated. Therefore, based on the spectrum analysis of signals in [33], the spectral trends of the smoothed OFDM signal can be accurately demonstrated by time-domain derivatives of the smoothed OFDM symbols. Furthermore, a compact upper bound of power leakage of the smooth signal is derived.

It is known that all derivatives of the OFDM signal $s(t)$ exist except at the points between adjacent OFDM symbols. Meanwhile, except these points, the smooth signal also possesses derivatives of all orders due to the existence of all derivatives of the basis function $f^{(\tilde{v})}(n)$. Then, we make the assumption that at the adjacent points, the first (*V*-1) derivatives of the smoothed OFDM signal $\overline{s}(t)$ is continuous, and the *V*th-order derivative $\overline{s}^{(V)}(t)$ of $\overline{s}(t)$ has finite amplitude discontinuities. In addition, suppose that all derivatives of the smoothed OFDM signal approach zeros as $t \to \pm\infty$. Based on (5), we obtain

$$\mathcal{F}\left(\overline{s}(t)\right) = \frac{1}{(j2\pi f)^{V-1}} \int_{-\infty}^{+\infty} \overline{s}^{(V-1)}(t) e^{-j2\pi ft} dt. \quad (31)$$

Furthermore, since $\overline{s}^{(V)}(t)$ has finite amplitude discontinuities, performing an integration by parts on the integrals in (31), by setting $u = \overline{s}^{(V-1)}(t)$ and $dv = e^{-j2\pi ft}dt$, we have

$$\mathcal{F}\left(\overline{s}(t)\right) = \frac{1}{(j2\pi f)^{V-1}}$$
$$\left( \frac{\overline{s}^{(V-1)}(t)e^{-j2\pi ft}}{-j2\pi f}\bigg|_{t=-\infty}^{t=+\infty} - \int_{-\infty}^{+\infty} \frac{s^{(V)}(t)e^{-j2\pi ft}}{-j2\pi f}dt \right)$$
$$= \frac{1}{(j2\pi f)^{V}} \int_{-\infty}^{+\infty} \overline{s}^{(V)}(t)e^{-j2\pi ft}dt.$$

There are finite amplitude discontinuities in $\overline{s}^{(V)}(t)$ at the adjacent points. Thus, $\overline{s}^{(V)}(t)$ can be written as

$$\overline{s}^{(V)}(t) = \sum_{i=-\infty}^{+\infty} \overline{y}_i^{(V)}\left(t - i(T_s + T_{cp})\right). \quad (32)$$

It is inferred by (32) that the *V*th derivative $\overline{y}_i^{(V)}(t)$ of the smoothed OFDM symbol is windowed by the rectangular function $R(t)$. Therefore, based on (1)-(3), the spectrum of the smoothed OFDM signal $\overline{s}(t)$ can be expressed by

$$\mathcal{F}\left(\overline{s}(t)\right) = \frac{\mathcal{F}\left(\overline{s}^{(V)}(t)\right)}{(j2\pi f)^{V}} = \frac{E\left\{\mathcal{F}\left(\overline{y}_i^{(V)}(t)\right)\right\}}{(j2\pi f)^{V}}. \quad (33)$$

Eq. (33) denotes that the spectrum of the smoothed OFDM signal can be estimated by the statistical average of the spectrum of the high-order derivatives of the OFDM symbols. From (3), the closed-form spectrum of $\overline{y}_i^{(V)}(t)$ can be calculated by

$$\mathcal{F}\left(\overline{y}_i^{(V)}(t)\right) = \frac{(j2\pi f)^{-V}}{T_s + T_{cp}} \int_{-T_{cp}}^{T_s} \overline{y}_i^{(V)}(t) R(t) e^{-j2\pi ft} dt$$
$$= \frac{(j2\pi f)^{-V}}{T_s + T_{cp}} \int_{-T_{cp}}^{T_s} \left(y_i(t) + w_i(t)\right)^{(V)} R(t) e^{-j2\pi ft} dt$$
$$= (j2\pi f)^{-V} \sum_{k_m \in \mathcal{K}_0} (\frac{j2\pi}{T_s}k_m)^V \left(X_{i,k_m} + W_{i,k_m}\right)$$
$$\text{sinc}\left(f_m(1+T_{cp}/T_s)\right)e^{j\pi f_m\left(1-\frac{T_{cp}}{T_s}\right)}.$$
$$\quad (34)$$

As illustrated by (34), the spectrum of $\overline{y}_i(t)$ can be expressed by the superposition of the spectrum of the *V*th-order derivative $y_i^{(V)}(t)$ of original OFDM symbol and the spectrum of the *V*th-order derivative $w_i^{(V)}(t)$ of the smooth signal divided by $(j2\pi f)^V$. It is denoted that the sidelobe of $\overline{s}(t)$, whose first $V$-1 derivatives are continuous, decays as $f^{-V-1}$. It is shown that $w_i(t)$ improves the time-domain continuity between adjacent OFDM symbols so that $\bar{s}(t)$ is with excellent out-of-band spectrum.

Finally, based on (33), the PSD of $\bar{s}(t)$ is given by

$$\bar{\psi}(f) = (T_s + T_{cp})E\left\{\left|\mathcal{F}\left(\bar{y}_i^{(V)}(t)\right)\right|^2\right\}$$

$$= \frac{T_s + T_{cp}}{(2\pi f)^{2V}} E\left\{\left|\sum_{k_m \in \mathcal{K}_0} (\frac{j2\pi}{T_s} k_m)^V \left(X_{i,k_m} + W_{i,k_m}\right)\right.\right.$$

$$\left.\left. \mathrm{sinc}\left(f_m(1+T_{cp}/T_s)\right)e^{j\pi f_m\left(1-\frac{T_{cp}}{T_s}\right)}\right|^2\right\}$$

$$= \frac{T_s + T_{cp}}{(T_s f)^{2V}} E\left\{\left|\sum_{k_m \in \mathcal{K}_0} k_m^V \left(X_{i,k_m} + W_{i,k_m}\right)\right.\right.$$

$$\left.\left. \mathrm{sinc}\left(f_m(1+T_{cp}/T_s)\right)e^{j\pi f_m\left(1-\frac{T_{cp}}{T_s}\right)}\right|^2\right\}. \quad (35)$$

where upper bound of the power leakage decaying is demonstrated. As indicated by (35), small power spectral sidelobes of $\bar{s}(t)$ roll off rapidly as at least $f^{-2V-2}$, which gives more compact spectrum roll-off than (5). Compared to the PSD of $s(t)$, it is also demonstrated that $\bar{s}(t)$ is with much deeper out-of-band spectrum.

According to the above analysis, the spectrum of $\bar{s}(t)$ is derived by the relationship between time-domain continuity and spectral trends. Contrary to NC-OFDM, TD-NC-OFDM gives a direct time-domain structure of the smoothed signal in terms of continuity of high-order derivatives. Thus, it is convenient to analyze the sidelobe decaying. Furthermore, Fig. 3 proves that the upper bound of (35) is a compact one for describing the power leakage decaying. We show that the sidelobe decaying is determined by the MDO, which indicates how many times the smoothed OFDM signal can be differentiated. As shown in Fig. 3, the power leakage of the smoothed OFDM signal by TD-NC-OFDM decays at least $f^{-2V-4}$ with the increasing of MDO $V$.

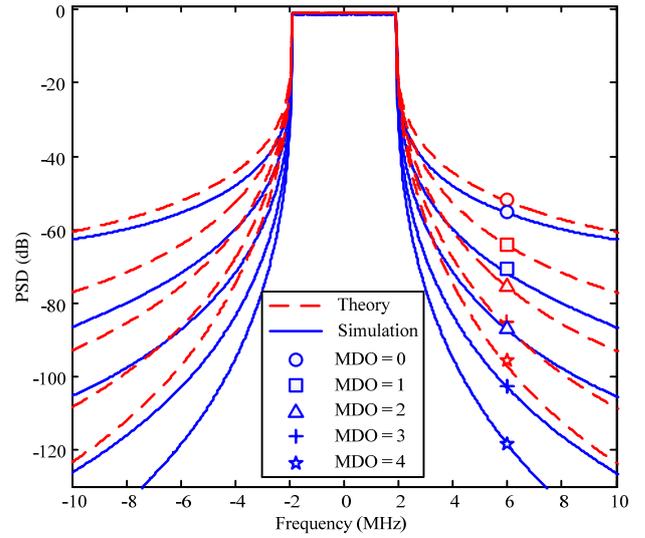

Fig. 3. PSD comparison between theory analysis and simulation results for NC-OFDM signals.

### B. Complexity Comparison

In NC-OFDM, its frequency-domain precoder consumes $2K^2$ complex multiplications and $2K^2$ complex additions as indicated by (7). After further reducing the scale of matrices, the complexity is reduced to $2VK+(V+1)(2V+1)$ complex multiplications, and $2VK+V(2V+1)$ complex additions by (20). Then, during the generation and overlapping of the smooth signal, $N(V+1)$ complex multiplications, and $N(V+1)$ complex additions are required in (21).

Assume that a complex addition is equivalent to two real additions; a complex multiplication to four real multiplications and two real additions; a real-complex multiplication to two real multiplications. The complexity of TD-NC-OFDM can be expressed as

$$\begin{aligned} C_{mul} &= 8VK + 4N(V+1) + 4(V+1)(2V+1) \\ C_{add} &= 8VK + 4N(V+1) + 2(2V+1)^2 \end{aligned} \quad (36)$$

where $C_{mul}$ and $C_{add}$ are the number of real multiplications and the number of real additions, respectively. Table I gives the complexity comparison between NC-OFDM and TD-NC-OFDM. Because of $VN \ll K^2$, the complexity of TD-NC-OFDM is much lower than that of NC-OFDM.

TABLE I. COMPLEXITY COMPARISON BETWEEN NC-OFDM AND TD-NC-OFDM

| Scheme | Real multiplication | Real addition |
|---|---|---|
| **NC-OFDM** | $8K^2$ | $8K^2-2K$ |
| **TN-NC-OFDM** | $4(V+1)N+8VK+$ $4(V+1)(2V+1)$ | $4(V+1)N+8VK$ $+2(2V+1)^2$ |

## C. SINR Analysis

A disadvantage of NC-OFDM is that the transmit signal will be interfered by the smooth signal. Thus, a measurement is required for denoting the interference into SNR loss. In this section, we investigate the SINR of NC-OFDM, based on mathematical analysis of the average power of the smooth signal.

In multipath channel with time-domain coefficients $h_{\tilde{l}}$ in the $\tilde{l}$ th path, the $i$th received time-domain OFDM symbol $r_i(t)$ is

$$r_i(t) = \sum_{\tilde{l}=1}^{\tilde{L}} h_{\tilde{l}} \overline{y}_i(t - \tau_{\tilde{l}}) + n_i(t) \quad (37)$$

where $\tau_{\tilde{l}}$ is the time delay in the $\tilde{l}$ th path, and $n_i(t)$ is AWGN noise with zero mean and variance $\sigma_n^2$. We assume that the length of CP is longer than or equal to the maximum channel time delay, thus the received SINR $\gamma_{SINR}$ can be expressed by

$$\gamma_{SINR} = \frac{\sum_{\tilde{l}=1}^{\tilde{L}} |h_{\tilde{l}}|^2 E\{\mathbf{y}_i^H \mathbf{y}_i\}}{\sigma_n^2 + \sum_{\tilde{l}=1}^{\tilde{L}} |h_{\tilde{l}}|^2 E\{\mathbf{w}_i^H \mathbf{w}_i\}}. \quad (38)$$

If the data vector $\mathbf{X}_i$ is uncorrelated, that is $E\{\mathbf{X}_i \mathbf{X}_i^H\} = \mathbf{I}_K$ and $E\{\mathbf{X}_{i-1} \mathbf{X}_i^H\} = \mathbf{0}_{K \times K}$, we can obtain $E\{\mathbf{X}_i^H \mathbf{X}_i\} = Tr\{E\{\mathbf{X}_i \mathbf{X}_i^H\}\} = K$. Thus, the average power of OFDM symbol vector $\mathbf{y}_i$ is

$$E\{\mathbf{y}_i^H \mathbf{y}_i\} = \frac{1}{N^2} E\{\mathbf{X}_i^H \mathbf{F}_f \mathbf{F}_f^H \mathbf{X}_i\} = E\{\mathbf{X}_i^H \mathbf{X}_i\} = \frac{K}{N}. \quad (39)$$

Then, the average power of the smooth signal is given by

$$\begin{aligned} E\{\mathbf{w}_i^H \mathbf{w}_i\} &= E\{(\mathbf{Q}_f \mathbf{b}_i)^H \mathbf{Q}_f \mathbf{b}_i\} \\ &= Tr\{E\{\mathbf{Q}_f \mathbf{b}_i \mathbf{b}_i^H \mathbf{Q}_f^H\}\} = \frac{2}{N}(V+1). \end{aligned} \quad (40)$$

The derivation of Eq. (40) is shown in Appendix C.

From (39) and (40), the power ratio between the transmit and the smooth signal is achieved by

$$\frac{E\{\mathbf{y}_i^H \mathbf{y}_i\}}{E\{\mathbf{w}_i^H \mathbf{w}_i\}} = \frac{K/N}{2(V+1)/N} = \frac{K}{2(V+1)}. \quad (41)$$

Consequently, the received SINR $\gamma_{SINR}$ is calculated by

$$\gamma_{SINR} = \frac{K/N}{\dfrac{\sigma_n^2}{\sum_{\tilde{l}=1}^{\tilde{L}} |h_{\tilde{l}}|^2} + \dfrac{2(V+1)}{N}}. \quad (42)$$

It is concluded by (42) that the received SINR is degraded as the MDO is increased. Fig. 4 shows some numerical results of the received SINR in simulation and its mathematical expression based on (42). The Rayleigh channel adopts the EVA channel model. Horizontal ordinate denotes the transmitted bit SNR, also known as $E_b/N_0$. Fig. 4 proves that the analysis of the received SINR based on (42) is close to the simulated SINR. We show in Fig.4 that SINR will approach constant when $E_b/N_0$ is high, since in this case the influence of noise can be neglected. Furthermore, we also show that SINR degradation will be more serious with the increase of MDO.

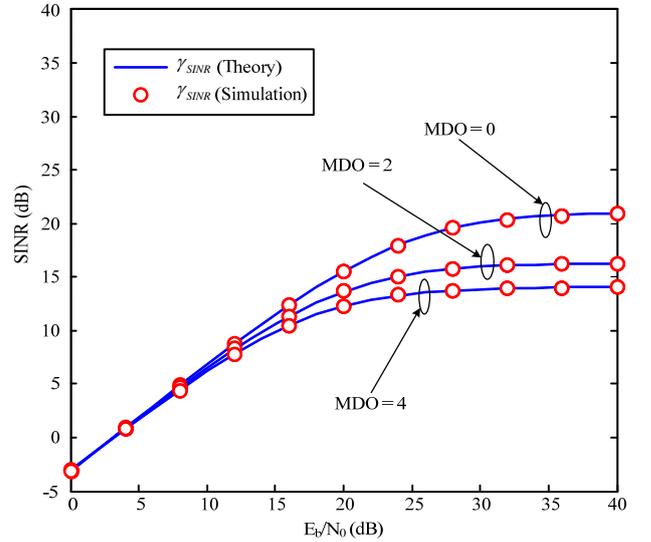

Fig. 4. Analysis and Simulations on SINR of NC-OFDM signals in EVA fading channel.

## VI. NUMERICAL RESULTS

In this section, we give simulation results to evaluate the sidelobe suppression, complexity, and BER performance of NC-OFDM (including TD-NC-OFDM). The simulation is performed for a baseband-equivalent OFDM system with 256 subcarriers, mapped onto the subcarrier index set $\{-128,\ldots,127\}$. Here, 16-QAM digital modulation is examined with symbol period $T_s = 1/15$ms, time-domain oversampling interval $T_{samp} = T_s / 2048$ and CP duration $T_{cp} = 144 T_{samp}$. To illustrate the sidelobe suppression effect, the PSD is evaluated by Welch's averaged periodogram method with a 2048-sample Hanning window and a 512-sample overlap after observing $10^5$ symbols. To investigate the error performance in the receiver, the signal is transmitted through EVA channel [36] with the maximum Doppler frequency offset 222Hz.

The PSD of NC-OFDM and TD-NC-OFDM transmit signals with different maximum derivative orders (MDOs) is shown in Fig. 5. It is depicted that TD-NC-OFDM has identical sidelobe suppression performance to NC-OFDM, which proves the equivalence between the two systems. As the MDO value increases, the sidelobe suppression performance is further improved. For example, at frequency 4MHz, the original OFDM signal has a PSD over 35dB and TD-NC-OFDM significantly reduces the PSD to approximately 70dB with $V$=2. When the MDO increases to 4, TD-NC-OFDM can reduce the PSD to approximately 97dB.

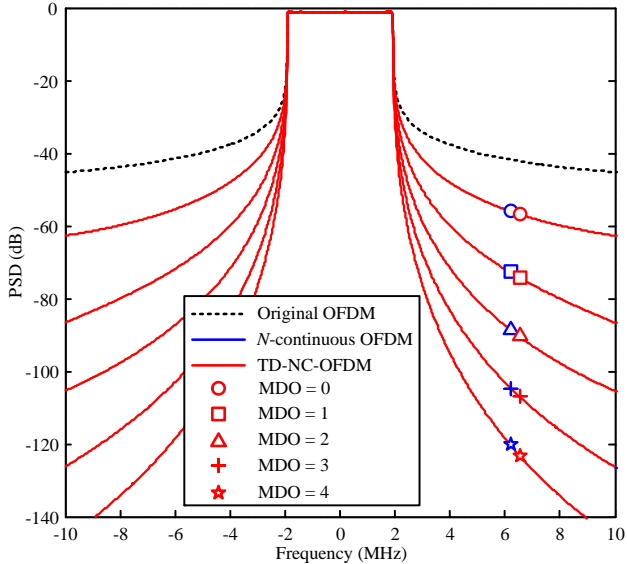

Fig. 5. PSD of NC-OFDM and TD-NC-OFDM transmit signals.

The computational complexity of NC-OFDM and TD-NC-OFDM are separately shown in Fig. 6 and Table II, in terms of real multiplications and additions, according to the analysis in section V-B. It is shown that TD-NC-OFDM is with dramatically reduced complexity over traditional NC-OFDM. For example, with $V$=2, the number of real multiplications and real additions in TD-NC-OFDM are only 5.5% of NC-OFDM. Therefore, TD-NC-OFDM has much lower complexity than NC-OFDM, while offering notable sidelobe suppression,.

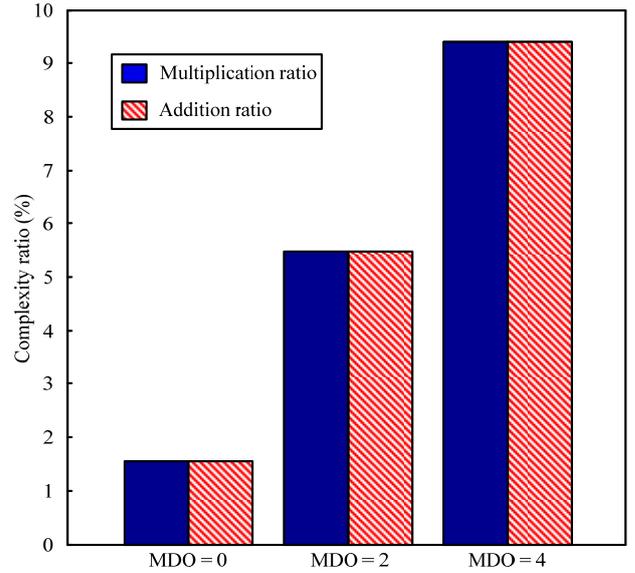

Fig. 6. Complexity ratio histogram of TD-NC-OFDM over NC-OFDM.

TABLE II. NUMERICAL COMPLEXITY COMPARISON BETWEEN NC-OFDM AND TD-NC-OFDM

| Scheme | | Real multiplications | Real additions |
|---|---|---|---|
| NC-OFDM ($V$=0, 2, 4) | | 524288 | 523776 |
| TD-NC-OFDM | $V$=0 | 8196 | 8196 |
| | $V$=2 | 28732 | 28722 |
| | $V$=4 | 49332 | 49314 |

In Fig. 7, we present simulation results for NC-OFDM and TD-NC-OFDM to illustrate the BER performance of varying MDOs in EVA channel. It is shown that TD-NC-OFDM has identical BER performance to NC-OFDM which also proves the equivalence between TD-NC-OFDM and NC-OFDM as before. Furthermore, Fig.7 shows that the system performance of the received signal is degraded as the MDO is increased. We show that the BER performance of NC-OFDM is consistent with the SINR analysis in section V-C.

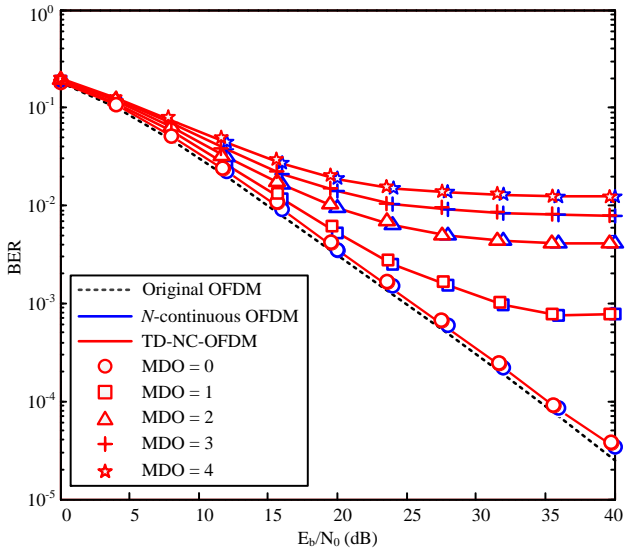

Fig. 7. BER of NC-OFDM and TD-NC-OFDM with varying MDOs in EVA channel.

## VII. CONCLUSION

In this paper, TD-NC-OFDM was proposed to provide appreciated system optimization in terms of complexity reduction in NC-OFDM. On one hand, the time-domain model was first proposed to make OFDM signal $N$-continuous by overlapping the smooth signal, as the linear combination of proposed basis set. Secondly, the smooth signal structure and the basis set were proved by the analyzed proposition. Thirdly, by reducing the scale of matrices, we designed a low-complexity TD-NC-OFDM transmitter. Finally, we proved the equivalence between TD-NC-OFDM and NC-OFDM.

On the other hand, the performance analysis of spectrum, complexity and SINR was considered. By the continuity criterion, we gave a closed-form spectrum expression and provided a compact upper bound of power spectrum sidelobes for TD-NC-OFDM. Then the complexity of TD-NC-OFDM was measured. Lastly, we analyzed the exact expression of the received SINR.

In general, we showed that TD-NC-OFDM can effectively suppress the sidelobe as well as NC-OFDM but with much lower complexity. In this case, TD-NC-OFDM is a promising implement way for conventional NC-OFDM.

## APPENDIX A

*Proof*: As indicated in [34], the reconstructed time-domain signal is expressed by

$$s(t) = \sum_{n=-\infty}^{+\infty} s(n) \frac{\sin\left(\pi(t-nT_{samp})/T_{samp}\right)}{\pi(t-nT_{samp})/T_{samp}}. \quad (43)$$

Differentiating the equation above in a $v$ time, we have

$$s^{(v)}(t) = \sum_{n=-\infty}^{+\infty} s(n) \left(\frac{\sin\left(\pi(t-nT_{samp})/T_{samp}\right)}{\pi(t-nT_{samp})/T_{samp}}\right)^{(v)}. \quad (44)$$

It is concluded by (44) that the $v$th-order derivatives of the continuous-time signal can be reconstructed by an interpolator, which linearly combines the $v$th-order derivative of *sinc* function. Meanwhile, the weighted coefficients of the derivatives of *sinc* function are signal values at sampling time.

## APPENDIX B

*Proof*: In a general multicarrier system, there may be multiple subbands. In the following, the multicarrier signal in one subband with the subcarrier index set $\mathcal{K}$ is considered. Similar to the expression of OFDM symbol in (1), the multicarrier symbol is also processed by the inverse discrete Fourier transformation (IDFT). Therefore, with the time-domain sampling interval $T_{samp}$ and CP duration $T_{cp}$, the $v$th-order derivative $y_i^{(v)}(n)$ of the $i$th discrete-time multicarrier symbol is given by

$$y_i^{(v)}(n) = \frac{1}{N} \sum_{k_m \in \mathcal{K}} (j2\pi \frac{k_m}{N})^v X_{i,k_m} e^{j2\pi \frac{k_m}{N} n}, \quad (45)$$

where $n \in \mathcal{N}$. Then, the spectral function $Y_i^{(v)}(k_m)$ of $y_i^{(v)}(n)$ is expressed by

$$Y_i^{(v)}(k_m) = \begin{cases} (j\frac{2\pi}{N} k_m)^v e^{-j\varphi k_m} X_{i,k_m}, & k_m \in \mathcal{K} \\ 0, & else \end{cases}, \quad (46)$$

which can be rewritten by

$$Y_i^{(v)}(k_m) = X_{i,k_m} F(k_m, v) \quad (47)$$

where the function $F(k_m, v)$ is defined by

$$F(k_m, v) = \begin{cases} (j\frac{2\pi}{N} k_m)^v e^{-j\varphi k_m}, & k_m \in \mathcal{K} \\ 0, & else \end{cases}. \quad (48)$$

The circular convolution theorem of discrete Fourier transformation (DFT) states that [34]

$$Y_i(k_m) F(k_m, v) \xrightleftharpoons[DFT]{IDFT} \tilde{y}_i(n) \circledast f(n, v). \quad (49)$$

Hence, the $v$th-order derivative $y_i^{(v)}(n)$ is also expressed by

$$y_i^{(v)}(n) = \tilde{y}_i(n) \circledast f(n, v) \quad (50)$$

where $f(n, v)$ is the IDFT of the frequency-domain $F(k_m, v)$

$$f(n,v) = \frac{1}{N} \sum_{k_m \in \mathcal{K}} (j2\pi \frac{k_m}{N})^v e^{-j\varphi k_m} e^{j2\pi \frac{k_m}{N} n}. \qquad (51)$$

Similar to the derivative theorem of the Fourier transform, $f(n,v)$ is the $v$th-order derivative of the function $f(n,0)$, which is given by

$$f(n,0) = \frac{1}{N} \sum_{k_m \in \mathcal{K}} e^{-j\varphi k_m} e^{j2\pi \frac{k_m}{N} n}. \qquad (52)$$

For convenience, the following $f(n,v)$ is denoted as $f^{(v)}(n)$.

Then, $\tilde{y}_i(n)$ denotes the symbol sample value by performing $K$-point IDFT for the $K$ data $X_{i,k_m}$. It can also be expressed by the sample of $y_i(n)$ at non-oversampling point

$$\begin{aligned}\tilde{y}_i(n) &= y_i(n) \sum_{m=0}^{K-1} \delta(n - Jm) \\ &= \sum_{m=0}^{K-1} y_i(Jm) \delta(n - Jm) \\ &= \sum_{\substack{m=0 \\ r=Jm}}^{K-1} y_i(r) \delta(n - r)\end{aligned} \qquad (53)$$

where $r = Jm$ with the oversampling factor $J = N/K$.

Therefore, the $v$th-order derivative $y_i^{(v)}(n)$ in (50) is rewritten by

$$\begin{aligned}y_i^{(v)}(n) &= \sum_{\substack{m=0 \\ r=Jm}}^{K-1} y_i(r) \delta(n-r) \circledast f^{(v)}(n) \\ &= \sum_{\substack{m=0 \\ r=Jm}}^{K-1} y_i(r) f^{(v)}((n-r)_N) \\ &= \tilde{y}_i(n) \circledast f^{(v)}(n).\end{aligned} \qquad (54)$$

When the number of subbands is larger than one, the $v$th-order derivative of multicarrier symbol in the whole band is the superposition of the $v$th-order derivative of multicarrier symbol per subband.

## APPENDIX C

Because the information data vector $\mathbf{X}_i$ is unrelated, we also obtain $E\{\mathbf{y}_i \mathbf{y}_i^H\} = \frac{1}{N^2} E\{\mathbf{F}_f^H \mathbf{X}_i \mathbf{X}_i^H \mathbf{F}_f\} = \frac{1}{N^2} \mathbf{F}_f^H \mathbf{F}_f$ and $E\{\mathbf{y}_{i-1} \mathbf{y}_i^H\} = \frac{1}{N^2} E\{\mathbf{F}_f^H \mathbf{X}_{i-1} \mathbf{X}_i^H \mathbf{F}_f\} = \mathbf{0}_{N \times N}$. Then, $\overline{\mathbf{y}}_i$ also satisfies $E\{\overline{\mathbf{y}}_i \overline{\mathbf{y}}_i^H\} = \frac{1}{N^2} \mathbf{F}_f^H \mathbf{F}_f$. Assuming $E\{\overline{\mathbf{y}}_{i-1} \overline{\mathbf{y}}_{i-1}^H\} = \frac{1}{N^2} \mathbf{F}_f^H \mathbf{F}_f$, based on (23), (25) and (26), we have

$$\begin{aligned}E\{\overline{\mathbf{y}}_i \overline{\mathbf{y}}_i^H\} &= E\{(\mathbf{y}_i + \mathbf{Q}_f \mathbf{b}_i)(\mathbf{y}_i + \mathbf{Q}_f \mathbf{b}_i)^H\} \\ &= E\{\mathbf{y}_i \mathbf{y}_i^H\} + \mathbf{Q}_f \mathbf{P}_f^{-1} \mathbf{P}_1 E\{\overline{\mathbf{y}}_{i-1} \overline{\mathbf{y}}_{i-1}^H\} (\mathbf{Q}_f \mathbf{P}_f^{-1} \mathbf{P}_1)^H \\ &\quad - \mathbf{Q}_f \mathbf{P}_f^{-1} \mathbf{P}_2 E\{\mathbf{X}_i \mathbf{X}_i^H\} (\mathbf{Q}_f \mathbf{P}_f^{-1} \mathbf{P}_2)^H \\ &= E\{\mathbf{y}_i \mathbf{y}_i^H\} + \frac{1}{N^2} \mathbf{Q}_f \mathbf{P}_f^{-1} \mathbf{B}_2 \mathbf{B}_2^H (\mathbf{P}_f^{-1})^H \mathbf{Q}_f^H \\ &\quad - \frac{1}{N^2} \mathbf{Q}_f \mathbf{P}_f^{-1} \mathbf{B}_2 \mathbf{B}_2^H (\mathbf{P}_f^{-1})^H \mathbf{Q}_f^H \\ &= \frac{1}{N^2} \mathbf{F}_f^H \mathbf{F}_f\end{aligned}$$

Moreover, the first smoothed OFDM signal is initialized by $\overline{\mathbf{y}}_0 = \mathbf{y}_0$, thus, $E\{\overline{\mathbf{y}}_0 \overline{\mathbf{y}}_0^H\} = \frac{1}{N^2} \mathbf{F}_f^H \mathbf{F}_f$. It is induced that all smoothed OFDM symbols satisfy

$$E\{\overline{\mathbf{y}}_i \overline{\mathbf{y}}_i^H\} = \frac{1}{N^2} \mathbf{F}_f^H \mathbf{F}_f. \qquad (55)$$

In the following, the average power of the smooth signal is derived. Based on (23) and (40), term $E\{\mathbf{Q}_f \mathbf{b}_i \mathbf{b}_i^H \mathbf{Q}_f^H\}$ is expressed by

$$\begin{aligned}E\{\mathbf{Q}_f \mathbf{b}_i \mathbf{b}_i^H \mathbf{Q}_f^H\} &= E\{\mathbf{Q}_f (\mathbf{P}_f^{-1} \mathbf{P}_1 \overline{\mathbf{y}}_{i-1} - \mathbf{P}_f^{-1} \mathbf{P}_2 \mathbf{X}_i) \\ &\quad (\mathbf{P}_f^{-1} \mathbf{P}_1 \overline{\mathbf{y}}_{i-1} - \mathbf{P}_f^{-1} \mathbf{P}_2 \mathbf{X}_i)^H \mathbf{Q}_f^H\} \\ &= E\{\mathbf{Q}_f \mathbf{P}_f^{-1} \mathbf{P}_1 \overline{\mathbf{y}}_{i-1} \overline{\mathbf{y}}_{i-1}^H \mathbf{P}_1^H (\mathbf{P}_f^{-1})^H \mathbf{Q}_f^H\} \\ &\quad + E\{\mathbf{Q}_f \mathbf{P}_f^{-1} \mathbf{P}_2 \mathbf{X}_i \mathbf{X}_i^H \mathbf{P}_2^H (\mathbf{P}_f^{-1})^H \mathbf{Q}_f^H\}.\end{aligned} \qquad (56)$$

Then, substituting (24)-(27) into the first term and the second term in (56), respectively, we obtain

$$\begin{aligned}E\{\mathbf{Q}_f \mathbf{P}_f^{-1} \mathbf{P}_1 \overline{\mathbf{y}}_{i-1} \overline{\mathbf{y}}_{i-1}^H \mathbf{P}_1^H (\mathbf{P}_f^{-1})^H \mathbf{Q}_f^H\} \\ = \frac{1}{N^4} \mathbf{Q}_f \mathbf{P}_f^{-1} \mathbf{B}_2 \mathbf{F}_f \mathbf{F}_f^H \mathbf{F}_f \mathbf{F}_f^H \mathbf{B}_2^H (\mathbf{P}_f^{-1})^H \mathbf{Q}_f^H \\ = \frac{1}{N^2} \mathbf{Q}_f \mathbf{P}_f^{-1} \mathbf{B}_2 \mathbf{B}_2^H (\mathbf{P}_f^{-1})^H \mathbf{Q}_f^H\end{aligned} \qquad (57)$$

and

$$\begin{aligned}E\{\mathbf{Q}_f \mathbf{P}_f^{-1} \mathbf{P}_2 \mathbf{X}_i \mathbf{X}_i^H \mathbf{P}_2^H (\mathbf{P}_f^{-1})^H \mathbf{Q}_f^H\} \\ = \frac{1}{N^2} E\{\mathbf{Q}_f \mathbf{P}_f^{-1} \mathbf{B}_2 \mathbf{\Phi} \mathbf{X}_i \mathbf{X}_{i-1}^H \mathbf{\Phi}^H \mathbf{B}_2^H (\mathbf{P}_f^{-1})^H \mathbf{Q}_f^H\} \\ = \frac{1}{N^2} \mathbf{Q}_f \mathbf{P}_f^{-1} \mathbf{B}_2 \mathbf{B}_2^H (\mathbf{P}_f^{-1})^H \mathbf{Q}_f^H.\end{aligned} \qquad (58)$$

Based on (57) and (58), the trace of $E\{\mathbf{Q}_f \mathbf{b}_i \mathbf{b}_i^H \mathbf{Q}_f^H\}$ is simplified as

$$Tr\left\{E\left\{\mathbf{Q}_f\mathbf{b}_i\mathbf{b}_i^H\mathbf{Q}_f^H\right\}\right\}$$
$$=\frac{2}{N^2}Tr\left\{\mathbf{Q}_f\mathbf{P}_f^{-1}\mathbf{B}_2\mathbf{B}_2^H(\mathbf{P}_f^{-1})^H\mathbf{Q}_f^H\right\} \quad (59)$$
$$=\frac{2}{N^2}Tr\left\{\mathbf{P}_f^{-1}\mathbf{B}_2\mathbf{B}_2^H(\mathbf{P}_f^{-1})^H\mathbf{Q}_f^H\mathbf{Q}_f\right\}$$
$$=\frac{2}{N}Tr\left\{(\mathbf{B}_2\mathbf{B}_2^T)^{-1}\mathbf{B}_2\mathbf{B}_2^H\left((\mathbf{B}_2\mathbf{B}_2^T)^{-1}\right)^H(\mathbf{B}_2\mathbf{B}_2^H)^T\right\}.$$

To calculate Eq. (59), the matrices operations $(\mathbf{B}_2\mathbf{B}_2^T)^{-1}\mathbf{B}_2\mathbf{B}_2^H\left((\mathbf{B}_2\mathbf{B}_2^T)^{-1}\right)^H(\mathbf{B}_2\mathbf{B}_2^H)^T$ are analyzed as follows.

Assuming $\mathbf{x}_v=[x_{v,1},x_{v,2},\cdots,x_{v,V+1}]^T$ is the eigenvector of matrix $\mathbf{B}_2\mathbf{B}_2^T$ corresponding to the eigenvalue $\lambda_v$, we have

$$\mathbf{B}_2\mathbf{B}_2^T\mathbf{x}_v=\lambda_v\mathbf{x}_v \quad (60)$$

where the element of $\mathbf{B}_2\mathbf{B}_2^T\mathbf{x}_v$ is

$$\left\{\mathbf{B}_2\mathbf{B}_2^T\mathbf{x}_v\right\}_v=\sum_{\bar{m}=1}^{V+1}\sum_{k=1}^{K}b_{vk}b_{\bar{m}k}x_{v,\bar{m}}. \quad (61)$$

Because the elements of $\mathbf{B}_2$ are separately pure imaginary and real in odd and even row, the element of $\mathbf{B}_2\mathbf{B}_2^H\mathbf{x}_v$ follows

$$\left\{\mathbf{B}_2\mathbf{B}_2^H\mathbf{x}_v\right\}_v=\sum_{\bar{m}=1}^{V+1}\sum_{k=1}^{K}b_{vk}b_{\bar{m}k}^*x_{v,\bar{m}}$$
$$=\begin{cases}\sum_{\bar{m}=1}^{V+1}\sum_{k=1}^{K}b_{vk}b_{\bar{m}k}x_{v,\bar{m}}, & v\text{ is even}\\ -\sum_{\bar{m}=1}^{V+1}\sum_{k=1}^{K}b_{vk}b_{\bar{m}k}x_{v,\bar{m}}, & v\text{ is odd}\end{cases}. \quad (62)$$

Thus, from (60) to (62), the eigenvalues of $\mathbf{B}_2\mathbf{B}_2^H$ follow

$$\mathbf{B}_2\mathbf{B}_2^H\mathbf{x}_v=\begin{cases}\lambda_v\mathbf{x}_v, & v\text{ is even}\\ -\lambda_v\mathbf{x}_v, & v\text{ is odd}\end{cases}. \quad (63)$$

$\mathbf{B}_2\mathbf{B}_2^T$ is a symmetric matrix, thus, there is an unitary matrix $\mathbf{U}$ satisfying [37]

$$\mathbf{B}_2\mathbf{B}_2^T=\mathbf{U}\mathbf{\Lambda}\mathbf{U}^{-1}=\mathbf{U}\mathbf{\Lambda}\mathbf{U}^H \quad (64)$$

where $\mathbf{\Lambda}=diag(\lambda_0\quad\lambda_1\quad\cdots\quad\lambda_V)$. Then, it is concluded from (63) and (64) that

$$\mathbf{B}_2\mathbf{B}_2^H=\mathbf{U}\mathbf{\Lambda}_1\mathbf{U}^H \quad (65)$$

where $\mathbf{\Lambda}_1=diag(\lambda_0\quad-\lambda_1\quad\cdots\quad(-1)^V\lambda_V)$. Moreover, the diagonal matrices $\mathbf{\Lambda}$ and $\mathbf{\Lambda}_1$ follow

$$\mathbf{\Lambda}^{-1}\mathbf{\Lambda}_1=diag(1\quad-1\quad\cdots\quad(-1)^V). \quad (66)$$

Substituting (64)-(66) into the matrices operations $(\mathbf{B}_2\mathbf{B}_2^T)^{-1}\mathbf{B}_2\mathbf{B}_2^H[(\mathbf{B}_2\mathbf{B}_2^T)^{-1}]^H(\mathbf{B}_2\mathbf{B}_2^H)^T$, we can obtain

$$(\mathbf{B}_2\mathbf{B}_2^T)^{-1}\mathbf{B}_2\mathbf{B}_2^H\left((\mathbf{B}_2\mathbf{B}_2^T)^{-1}\right)^H(\mathbf{B}_2\mathbf{B}_2^H)^T$$
$$=(\mathbf{U}\mathbf{\Lambda}\mathbf{U}^H)^{-1}\mathbf{U}\mathbf{\Lambda}_1\mathbf{U}^H\left((\mathbf{U}\mathbf{\Lambda}\mathbf{U}^H)^{-1}\right)^H(\mathbf{U}\mathbf{\Lambda}_1\mathbf{U}^H)^T \quad (67)$$
$$=\mathbf{U}\mathbf{\Lambda}^{-1}\mathbf{\Lambda}_1\mathbf{\Lambda}^{-1}\mathbf{\Lambda}_1\mathbf{U}^T$$
$$=\mathbf{I}_{V+1}.$$

Finally, the trace in (59), that is the average power of the smooth signal in time domain, is achieved by

$$Tr\{E\{\mathbf{Q}_f\mathbf{b}_i\mathbf{b}_i^H\mathbf{Q}_f^H\}\}=\frac{2}{N}Tr\{\mathbf{I}_{V+1}\}=\frac{2(V+1)}{N}. \quad (68)$$

ACKNOWLEDGMENT (HEADING 5)